\documentclass{iaus}
\usepackage{graphicx}

\def\aj{\textit{AJ}}               
           
\def\apj{\textit{ApJ}}   
\def\apjl{\textit{ApJ}}

\def\aap{\textit{A\&A}}            
\def\mnras{\textit{MNRAS}}

\title[The Tully-Fisher relation: evolution with redshift and environment] 
{The Tully-Fisher relation: \\ evolution with redshift and environment}

\author[Alfonso Arag\'on-Salamanca]   %% give here short author list %%
{Alfonso Arag\'on-Salamanca}

\affiliation{
School of Physics and Astronomy,
  University of Nottingham, University Park, Nottingham, NG7 2RD, UK.
 \break email: alfonso.aragon@nottingham.ac.uk}

\pubyear{2006}
\volume{235}  %% insert here IAU Symposium No.
\pagerange{1--4}
\date{?? and in revised form ??}
\jname{Galaxy Evolution across the Hubble Time}
\editors{F.\ Combes \& J.\ Palous, eds.}
\begin{document}

\maketitle

\begin{abstract}

The Tully-Fisher Relation (TFR) links two fundamental properties of disk 
galaxies: their luminosity and their rotation velocity (mass). The pioneering
work of Vogt et al.\ in the 1990's showed that it is possible to study the TFR
for spiral galaxies at considerable look-back-times, and use it as a powerful 
probe of their evolution. In recent years, several groups have studied the TFR
for galaxies in different environments reaching redshifts beyond one.  In this
brief  review I summarise the main results of some of  these studies and their
consequences for our understanding of the formation and evolution of disk
galaxies. Particular emphasis is placed on the possible environment-driven
differences in the behaviour of the TFR for field and cluster galaxies.

\end{abstract}

\firstsection % if your document starts with a section,
\section{Introduction}

The Tully-Fisher relation (TFR; Tully \& Fisher 1977) is a strong correlation 
between the luminosity and the maximum rotation velocity of spiral galaxies. 
Although its physical origin is not completely understood in detail, the TFR
has proved to be a very useful tool to study the evolution of the fundamental
properties of spiral galaxies. Vogt et al.\ (1996) showed  that using large
telescopes with efficient spectrographs it is possible to study the TFR
reaching $z\sim1$. Following that work, several groups have  carried out
similar studies for sizeable samples of disk galaxies over a wide range in
redshift and environment. 

\vspace{-0.3cm}

\section{Evolution of the Tully-Fisher Relation of field galaxies}

The optical TFR of field galaxies has been studied using slit spectroscopy 
obtained at the VLT, Subaru and Keck telescopes by several groups including
Vogt et al.\ (1996), Ziegler, et al.\ (2002), B{\"o}hm et al.\ (2004),  
Bamford et al.\ (2006), Nakamura et al.\ (2006) and Weiner et al.\ (2006). 
These studies find that at a fixed rotation velocity,  spiral galaxies appear
to be  $\sim0.5$--$1.5\,$mag brighter at $z\sim1$ than at  $z\sim0$ in
rest-frame $B$. If interpreted as  pure luminosity evolution, this brightening
implies a factor of $\sim 3$ increase is star-formation rate between $z\sim0$
and $z\sim1$  (Bamford et al.\ 2006; Nakamura et al.\ 2006).  However, the
sample selection and the exact details of how the comparison is made between
samples at  different redshifts have very severe effects on the measured
evolution (cf.\ Vogt 2006, this symposium).  
B{\"o}hm et al.\ (2004) found a  correlation between the optical TFR residuals
and the rotation velocity $V_{\rm rot}$ and argued that the observed luminosity
evolution is mass dependent, in the sense that low-mass galaxies seem to
exhibit a much stronger rate of evolution than massive ones (but see Weiner et
al.\ 2006 for an opposite view). However, Bamford et al.\ (2006) demonstrated 
that such a correlation may exist purely due to an intrinsic coupling between
the $V_{\rm rot}$  scatter and the TFR residuals, acting in combination with
the selection effects inherent to magnitude-limited samples. Thus, decoupling
real evolution from observational artifacts if far from trivial. 

Studies at longer wavelengths (e.g.,  Conselice et al.\ 2005)  find very little
or no evolution in the near-infrared (or stellar mass) TFR of field spirals up
to $z\sim1$, suggesting that baryonic mass is accreted by disks along with dark
matter at $z<1$.

\vspace{-0.3cm}

\section{The effect of the cluster environment on the Tully-Fisher Relation}

Kannappan et al.\ (2002) found that    the deviations from the TFR of nearby
galaxies   are strongly correlated with  their colours and H$\alpha$ equivalent
widths, and thus with  their star formation properties and history. 
Correlations with the degree of photometric and kinematic  asymmetry were also
found, presumably linked with the dynamical interactions and history of the
galaxies. Since these effects are environmentally-dependent, it is
clear that the TFR and its evolution can be strongly influenced by the
environment. 

Vogt at al.\ (2004a,b,c) carried out a detailed study of the effect
of the cluster environment on the rotation curves and TFR of low redshift
galaxies. One of their results is that cluster and field star-forming spirals
follow essentially the same TFR, while non-starforming galaxies seem to be 
too faint for their rotation velocity. These spirals could be on their way to
becoming S0s (see \S4).   

Comparisons of the TFRs of cluster and field spirals at redshifts reaching
$z\sim1$ have been published by Milvang-Jensen et al.\ (2003),  Ziegler et al.\
(2003), Bamford et al.\ (2005), Nakamura et al.\ (2006) and Metevier et al.\
(2006). Although most of these studies do not find significant differences
between the TFRs of field and cluster  spirals, Milvang-Jensen et al.\ and
Bamford at al.\ reported some evidence  suggesting that cluster star-forming
spirals are brighter, at a fixed $V_{\rm rot}$, than their field counterparts
at the same redshift. Their interpretation is that field  spirals falling into
rich clusters undergo a (short?) period of enhanced star formation before
having their star formation quenched by the cluster  environment and
transforming, perhaps, into S0s. Nakamura et al.\ (2006) showed that even if
such processes are taking place, their effect  on the TFR of star-forming
cluster galaxies could be quite difficult to detect. They also demonstrated
that the sample selection and the details of how the analysis is actually
carried out could have a very strong effect on the conclusions of cluster-field
comparisons.   

\vspace{-0.3cm}

\section{The Tully-Fisher Relation of lenticular (S0) galaxies }

\begin{figure}
\begin{center}
\includegraphics[height=2.5in,width=2.5in,angle=0]{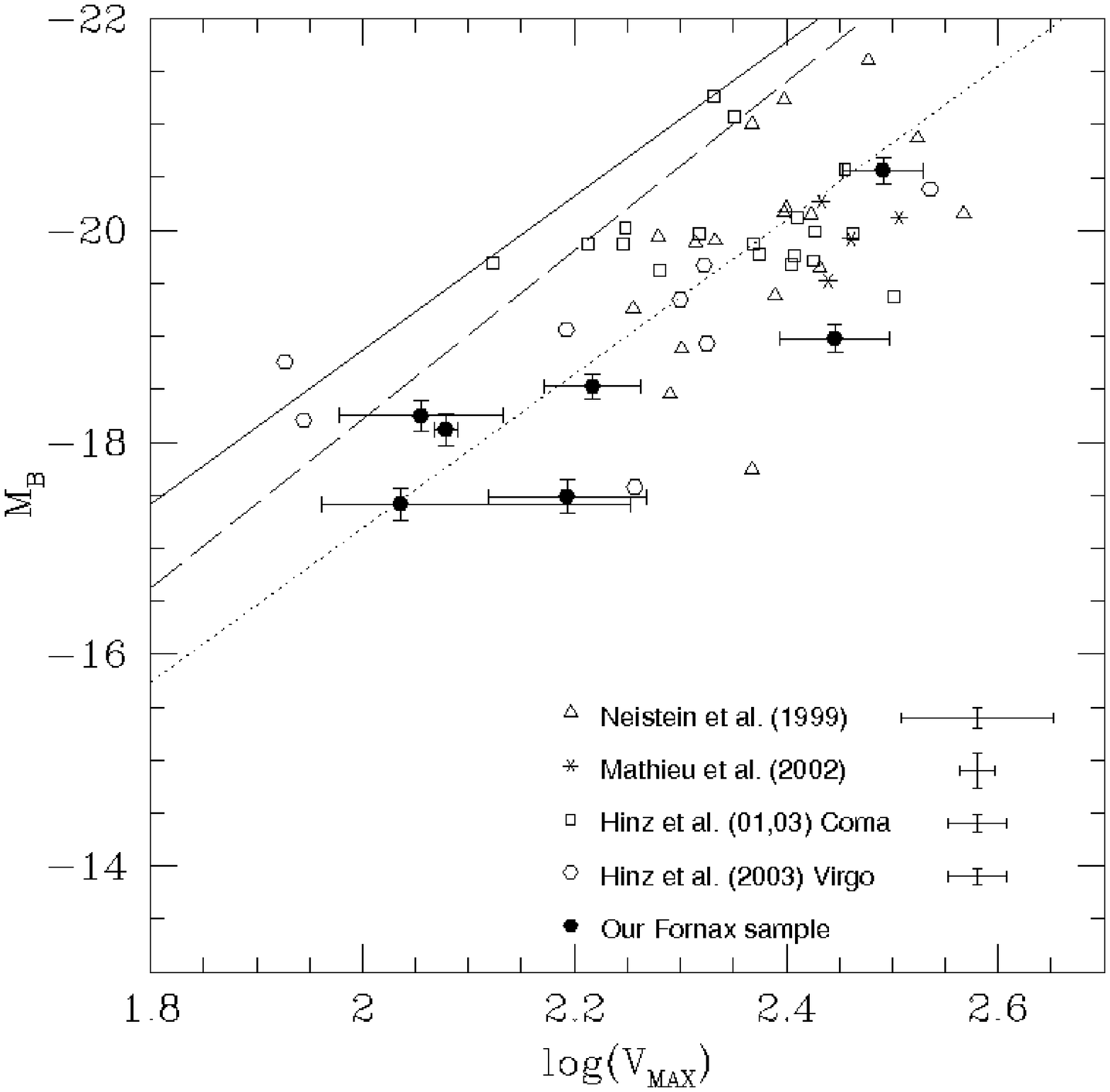}
\ \ \ \includegraphics[height=2.5in,width=2.5in,angle=0]{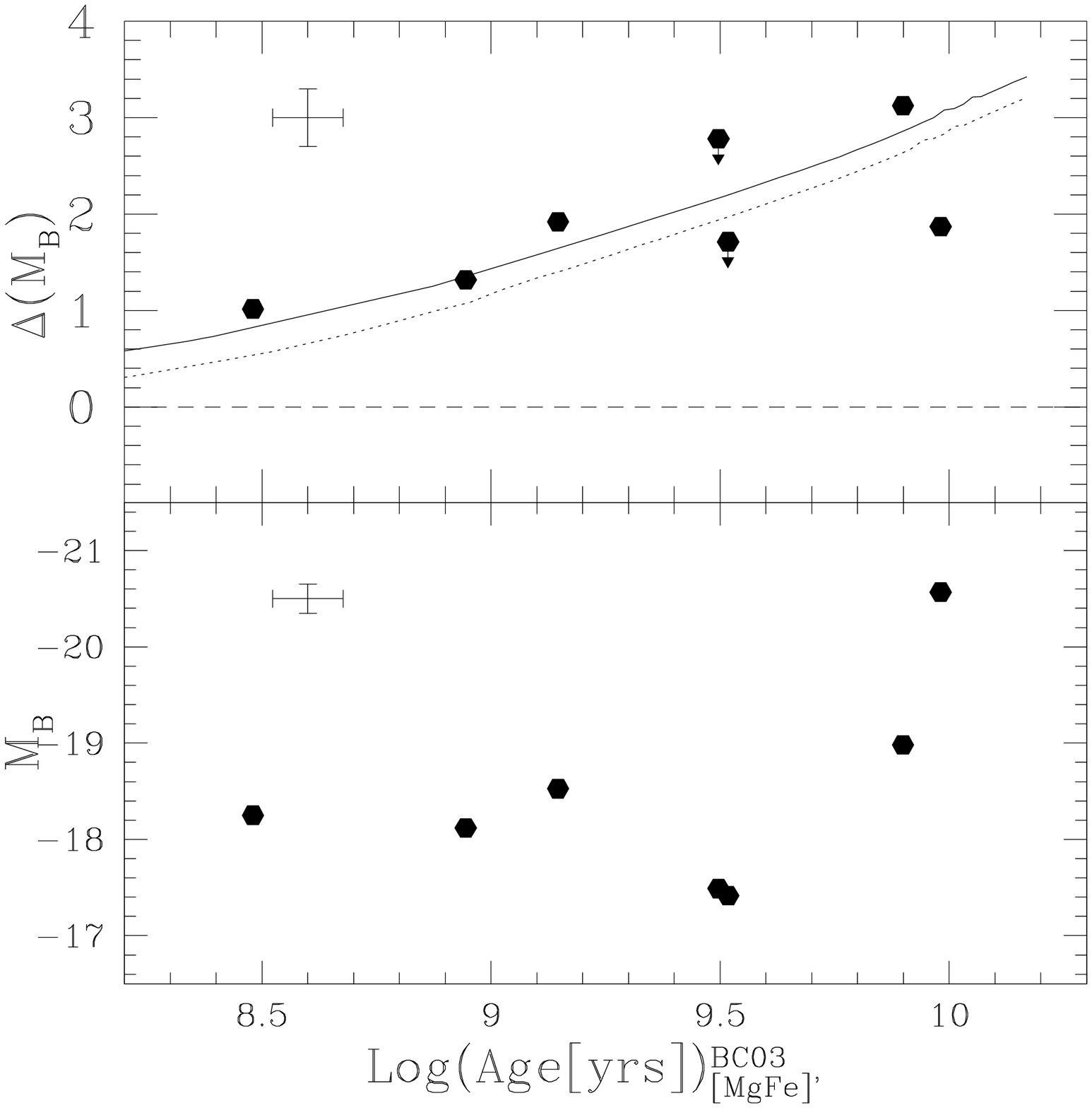}
\end{center}
\caption{
    (\textit{Left}) 
  $B$-band Tully--Fisher relation of S0 galaxies (datapoints). 
  Solid and dashed lines 
  represent two alternative determinations of the TFR of spiral galaxies,  
  while the dotted line is a fit to the S0 data.  
    (\textit{Right}) 
  The offset from the spiral TFR ($\Delta M_{\rm B}$)  and 
  absolute $B$ magnitudes ($M_{\rm B}$) versus the 
  Lick-indices determined ages of a sample of S0 galaxies 
  in the Fornax cluster. The solid and dotted lines 
  show population synthesis model predictions for 
  fading spiral galaxies. See  Bedregal et al.\ (2006a,b) for details.   
    }
\label{fig:fig1}
\end{figure}

In recent years several groups have extended the study of the TFR to S0
galaxies (see Bedregal et al.\ 2006a,b and references therein). Despite the 
considerable technical challenges involved in determining the true $V_{\rm
rot}$ from absorption line spectra of galaxies with significant non-rotational
support (see Mathieu et al.\ 2002), it seems clear now that S0s are, at
a given rotation velocity, fainter than spiral galaxies in the optical and 
near-infrared. The measured offset between the SO and spiral TFRs seems now to
be consistent with the hypothesis that S0s where once spirals that  ceased
forming stars. Indeed, the amount of fading implied by the  offset from the
spiral TFR of individual S0 galaxies seems to correlate with how long ago their
star formation ceased (see figure~\ref{fig:fig1} and Bedregal et al.\ 2006b).
Interestingly, a similar conclusion is reached from the properties of the
globular cluster systems of S0s and spirals  (Arag\'on-Salamanca et al.\ 2006).

\vspace{-0.3cm}

\section{Conclusions and future prospects}. 

Much has been learned so far on how the TFR of disk galaxies changes with
environment and redshift, but significant  uncertainties still remain. Progress
is being made  from larger galaxy samples like the ones now available in the
field (e.g., Weiner et al.\ 2006) and in clusters (e.g., the  ESO Distant
Cluster Survey, White et al.\ 2004). Based on these samples, new TFR  studies
comparing a range of field and cluster environments will soon become available
(Milvang-Jensen et al., Bamford et al., in preparation). Integral Field
Spectroscopy (IFS) is adding another dimension to  our understanding of the TFR
and the physical causes of its scatter and evolution (cf. Flores
et al.\ 2006). IFS coupled with the magnification  provided by gravitational
lensing is helping us reaching larger redshifts and higher spatial resolutions,
although the samples are still small  (Swinbank et al.\ 2006). Extending the
redshift baseline requires the efficient near-infrared spectrographs
(Smith et al.\ 2004, van Starkenburg et al.\ 2006) now becoming available 
at large telescopes.

\begin{acknowledgments}
I thank S.\ Bamford, A.G.\ Bedregal, 
M.\ Merrifield, B.\ Milvang-Jensen, O.\ Nakamura, N.\ Arimoto, C.\ Ikuta
and L.\ Simard for allowing me to discuss here 
results obtained with their help.  

\vspace{-0.5cm}

\end{acknowledgments}

\begin{discussion}

\discuss{Vogt}{Would you care to speculate on the formation mechanisms for the
high luminosity end of the S0 distribution?}

\discuss{Arag\'on-Salamanca}{They are probably not formed by disk galaxy infall
at $z\sim0.5$. The more massive ones are probably in place quite early, but we
do not really have much observational evidence on how they formed. }

\discuss{Bureau}{What are the prospects/future offered by integral field
spectroscopy?}

\discuss{Arag\'on-Salamanca}{As the work of Flores et al.\ and  Swinbank et
al.\ shows, integral field spectroscopy can provide not only an ``integrated''
rotation curve, as long-slit spectroscopy does, but also detailed kinematical
information. This additional information allows us to find out, for instance, 
whether galaxies which deviate from the TFR do so because of straightforward
luminosity evolution, of because of peculiar/disturbed kinematics. Thus
we can not only measure evolution/deviations from the
TFR, but also get a handle on the physical causes. }  

\discuss{Sadler}{Just a comment -- the original TFR in the local universe was
built on radio observations of the 21cm HI line. Radio HI studies at $z>0.1$
are only just becoming possible, but in the next 5 years it should be possible 
to measure HI rotation curves out to $z\simeq0.5$ with the technology
prototypes currently being built for the Square Kilometre Array. As the
collecting area increases it should eventually be possible to extend these
studies for $z\simeq1$ and beyond. } 

\discuss{Arag\'on-Salamanca}{I agree. Although this review concentrates mainly
on optical and near-IR studies (the ones that today are able to reach high
enough redshifts for evolutionary studies), I believe radio studies have the
potential to contribute greatly in this field. Indeed, at low redshifts HI
observations are already telling us a lot about how the environment affects
galaxy properties. }

\end{discussion}
\end{document}